\documentclass[aps, prl, letterpaper,twocolumn,english, floatfix]{revtex4}
\usepackage[T1]{fontenc}
\usepackage[latin1]{inputenc}
\usepackage{float}
\usepackage{graphicx}

\makeatletter

\usepackage{floatflt}

\usepackage{babel}
\makeatother
\begin{document}

\title{Ordered Nano-Crystal Arrays Spontaneously Form in Films Evaporated
onto Nanopore Array Substrates}

\author{Niravun Pavenayotin, M. D. Stewart, Jr., James M. Valles, Jr.}

\affiliation{Department of Physics, Brown University}

\author{Aijun Yin, J. M. Xu}

\affiliation{Division of Engineering and Department of Physics, Brown University}

\begin{abstract}
We present SEM images of films thermally evaporated onto Anodic Aluminum
Oxide substrates that are patterned with a hexagonal array of 34 and
80 nm diameter holes spaced by 100 nm. Over a range of film thicknesses,
Pb and Sn films spontaneously self assemble into an array of nano-crystals
in registry with the underlying hole lattice. The development of the
arrays with thickness indicates that surface energies drive coalescing
grains to move over the holes. Materials that wet the substrate or
whose grains do not coalesce at these substrate temperatures do not
form arrays. We discuss some potential applications.
\end{abstract}
\maketitle
Many approaches to create nanoscale structures in metals and semiconductors
have been explored for the development of improved devices such as
higher density magnetic memory \cite{1,1too,1three} and the investigation
of fundamental scientific problems, such as the Kondo effect \cite{2}.
One approach has been to grow films on substrates with nanometer scale
strucutre \cite{3,4} that serve as templates. For example, superconducting
films deposited on carbon nanotubes form wires that are suitable for
investigations of one dimensional superconductiviy \cite{5}. Also,
Anodic Aluminum Oxide (AAO) substrates with hexagonal nanopore arrays
have been employed to pattern superconducting Nb films with nano holes
on the order of the superconducting coherence length for the study
of vortex ordering \cite{6}.

In the cases mentioned above, the deposited films assumed the structure
of their substrates. That is, the metal films on carbon nanotubes
are as one dimensional as the nanotubes \cite{5} and the films on
the hexagonal pore array develop a hexagonal hole pattern \cite{6}.
Here we show that the interaction between the films and the substrate
can lead to film structure that differs dramatically from the substrate
structure. Specifically, Pb and Sn films thermally evaporated onto
AAO substrates spontaneously form a hexagonal array of nano-crystals
that are in registry with the holes (i.e. the negative of the substrate).
The detailed evolution of this film structure with thickness and comparisons
with the growth of other metals on these substrates suggest that surface
energies drive the nano-crystal array formation.

\begin{figure}[H]
\center

\includegraphics[%
  width=0.90\columnwidth,
  keepaspectratio]{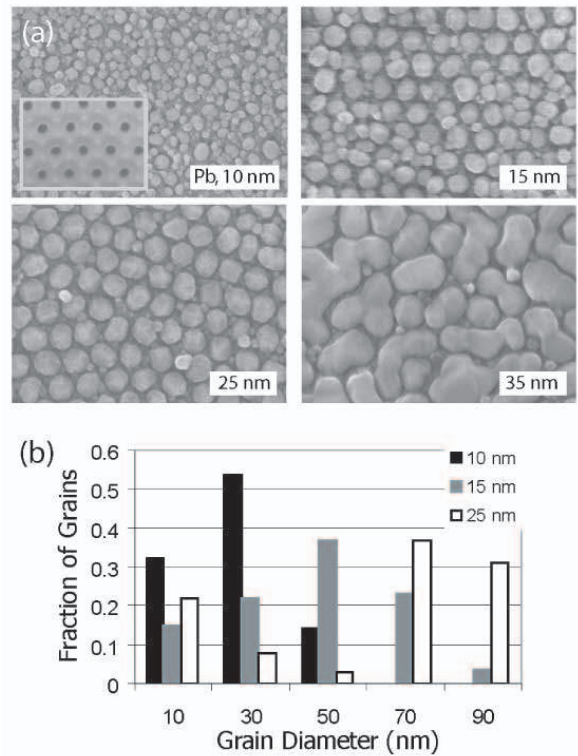}

\caption{a) Scanning electron microscope images of thermally evaporated Pb films
on AAO substrates with 34 nm diameter holes with 100nm center to center
spacing.  The nominal film thicknesses are indicated in each panel.
Inset: Image of a bare substrate. All of the images, including the
inset are presented at the same magnification. b)Histogram of the grain size distribution for each film
thickness. \label{cap:1}}
\end{figure}

The AAO substrates employed here were fabricated using established
procedures \cite{4,7,7too}. A Scanning Electron Microscopy image
of one of them is shown in the inset in Fig. \ref{cap:1}. The average
diameters of the holes in the arrays employed were $d_{H}=34$ and
80 nm and the standard deviation of the hole diameters was about 5\%.
The depth of the holes greatly exceeded the interhole spacing, $a=100$
nm. Elemental films (Pb, Sn, Au, Pd, Ge) were thermally evaporated
onto AAO substrates held at room temperature in a vacuum of approximately
$10^{-6}$ Torr. The quoted thicknesses correspond to the mass per
unit area deposited, as measured by a quartz crystal microbalance,
divided by the bulk density of the material. Each film shown in the
figures was made in a single evaporation step.

\begin{figure}[t]
\includegraphics[%
  width=2.25in,
  keepaspectratio]{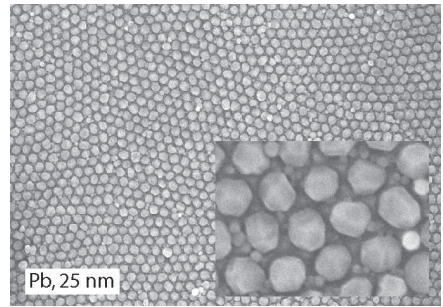}

\caption{SEM images at two magnifications of the 25 nm thick Pb film shown
in Fig. \ref{cap:1}. The grains in the images have the same center
to center spacing as the substrate. The inset provides a closeup of
the grains that reveals their faceted surfaces.\label{cap:2}}
\end{figure}

\begin{figure}[b]
\center

\includegraphics[%
  width=0.90\columnwidth,
  keepaspectratio]{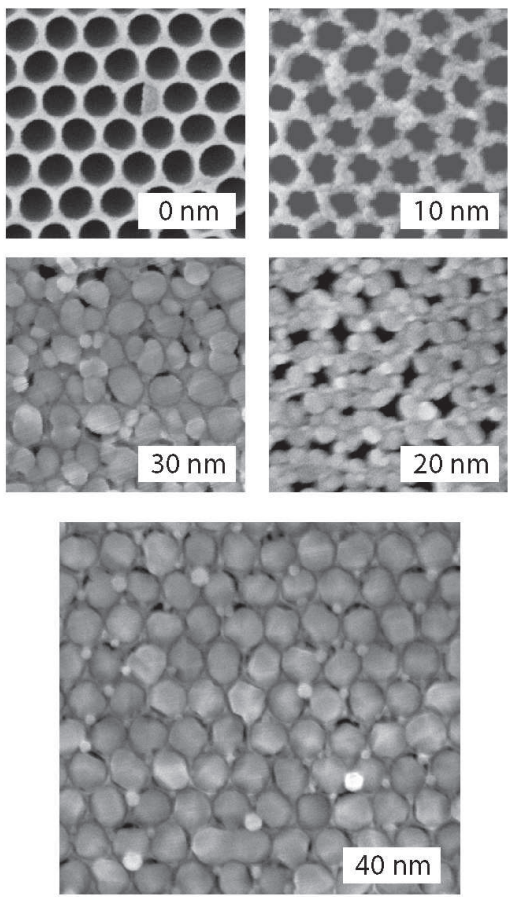}

\caption{SEM images of thermally evaporated Pb films on AAO substrates with
80 nm diameter holes and 100 nm center to center hole spacing. The
nominal film thicknesses are indicated in each panel. The 0 nm image
corresponds to a bare substrate. All of the images, including the
inset are presented at the same magnification.\label{cap:3}}
\end{figure}

The spontaneous formation of nano-crystals is demonstrated in Figs. \ref{cap:1}, \ref{cap:2}, and \ref{cap:3}.
The SEM images in Fig. \ref{cap:1} depict Pb films of different thicknesses
on the AAO substrates with 34 nm diameter holes. At the lowest thickness,
$t=10$ nm, the grains are randomly distributed and their size distribution
peaks at a diameter, $d_{G}$, near 30 nm, with few grains with $d_{G}>40$
nm. Interestingly, at this stage the holes in the substrate are almost
completely obscured by the grains. With increasing thickness to $t=15$
nm, larger grains appear. The grain size distribution develops a peak
near $d_{G}=50$ nm and extends to include grains 70 nm in diameter
and more. Simultaneously, the largest grains spontaneously assemble
into a hexagonal array. At $t=25$ nm, the grain diameter distribution
exhibits a peak at even higher diameters (centered on $d_{G}=80$
nm) and the hexagonal ordering is strongest. The spatial period of
the grain array coincides closely with that of the underlying lattice
of holes. This coincidence is brought out by the large area image
in Fig. \ref{cap:2}. Not only is the local ordering hexagonal, but
the grain arrangements exhibit domains and defects similar to those
in the nanopore arrangements. Finally, the magnified image in the
inset of Fig. \ref{cap:2} reveals that the large grains are faceted
indicating that they are single crystals.

Pb films deposited on AAO substrates with larger holes ($d_{H}=80$
nm) with the same spacing ($a=100$ nm) also spontaneously form an
ordered array of grains. The details of their morphological evolution
differs slightly (see Fig. \ref{cap:3}) from that described above
and in a manner that provides further insight into the grain array
formation process. At the lowest thicknesses, the Pb grains do not
obscure the holes, but rather, they `roughen' the hole edges. With
increasing film thickness, the grains grow and as they become comparable
to the size of the necks between holes, gradually obscure the holes.
Eventually, at $t=40$ nm the grains form an ordered array and assume
a relatively narrow size distribution similar to the results on the
substrate with 34 nm holes. The grain sizes (diameter of about 90
nm) and the film thickness ($t=40$ nm) necessary for the array formation,
however, are larger than on the 34 nm hole substrate. Finally, in
thicker films, grains covering neighboring holes start to coalesce
(not shown).

\begin{figure}[t]
\center

\includegraphics[%
  width=0.93\columnwidth, keepaspectratio]{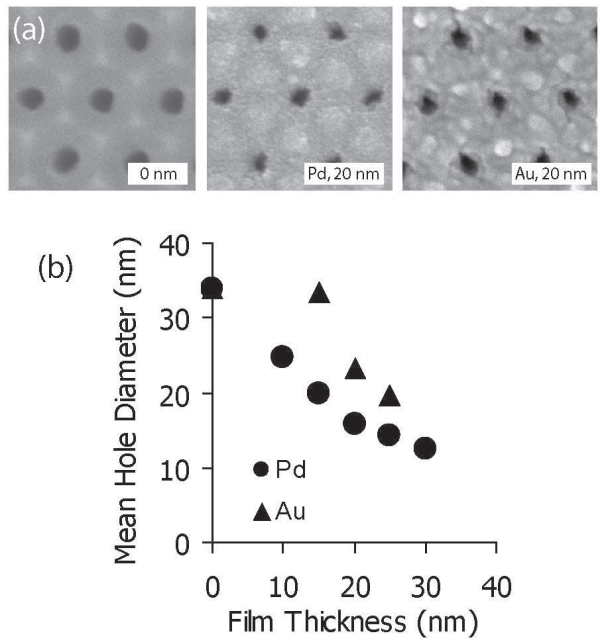} 

\caption{a) SEM images of a bare substrate and thermally evaporated 20 nm thick
Pb and Au films on AAO substrates with 34 nm diameter holes with 100
nm center to center spacing. All of the images, including the inset
are presented at the same magnification. b) Average
diameter of the hole openings as a function of film thickness for
Au and Pd.\label{cap:4}}
\end{figure}

We have also observed spontaneous nano-crystal array formation in
Sn films on 34 nm hole AAO substrates through a similar evolution
as shown in Fig. \ref{cap:1} for Pb. Again, the Sn grains grow with
film thickness and form over the holes at a thickness at which the
grains become comparable to the hole size. On the other hand, grain
arrays do not form for Au, Pd, and Ge films. As shown in Fig. \ref{cap:4},
Au and Pd films reproduce the substrate geometry, but gradually fill
in the holes as the film thickness increases. Unlike the Pb and Sn
films, these films either have grains that do not change their size
with film thickness (Au) or do not have grains at all (Ge, Pd).

There are two features exhibited during film growth that appear essential
for the formation of the ordered array of nano-crystals: 

\begin{enumerate}
\item The materials must not wet the substrate so that grains can form 
\item The temperature of the substrate, $T_{s}$, must be high enough that
the mobility of the deposited atoms, crystalline defects, and grain
boundaries is high enough to promote coalescence.
\end{enumerate}
Pb and Sn films, for which the ratios of the substrate temperature
to the elemental melting temperature, $T_{s}/T_{M}$, are lowest,
exhibit both of these features. Grains are clearly evident in their
SEM images indicating non-wetting of the substrate. The wide distribution
of grain diameters in the thinner films, the growth in the average
grain diameter with coverage and more directly, the observation of
grains caught in the process of combining (see $t=40$ nm film in
Fig. \ref{cap:1}) are consistent with grain growth through coalescence
and accretion of deposited atoms. During deposition, grains grow,
touch and coalesce. The randomness of the nucleation and coalescence
processes gives rise to the distribution of grain %

 sizes. By contrast, while Au films do form grains and thus do not
wet the substrate, their grain size distribution is comparatively
uniform and the average grain size does not change with thickness.
These characteristics indicate that grain coalescence does not occur
in them. Pd and Ge films do not form grains that are large enough
to be resolved in the SEM images and thus, appear to wet the substrate.

The fact that the Pb grains form in perfect registry with the holes
indicates that the grains find it energetically most favorable to
be off the substrate. A liquid drop model \cite{8} of the thin film
growth appears to be a sufficient description of the free energies
involved. The difference between the free energy of a grain on the
substrate and a grain of the same size and shape over a hole is approximately\begin{equation}
\Delta g=\left[\gamma_{i}a_{i}+\gamma_{m}a_{m}\right]-\left[\gamma_{m}\left(a_{i}+a_{m}\right)-\gamma_{s}a_{i}\right],\label{eq:deltag}\end{equation}
where $\gamma_{i}$ is the energy of the interface formed by the substrate
and the metal crystal facet in contact with it, $\gamma_{m}$ is the
surface energy of the metal crystal facet and $\gamma_{s}$ is the
substrate surface energy. The areas of the grain in contact with the
surface and not in contact with the surface are $a_{i}$ and $a_{m}$
respectively. The total area of a grain is $a_{i}+a_{m}$. For a grain
over a hole to be in the lowest free energy state, $\Delta g<0$,
or $\gamma_{i}>\gamma_{m}+\gamma_{s}$.

\begin{figure}
\begin{center}
\includegraphics[scale=0.35]{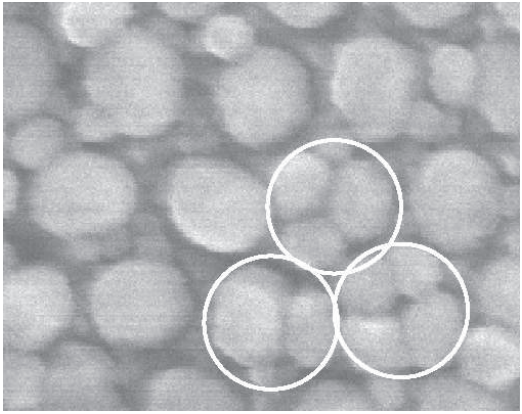}
\end{center}

\caption{Close up view of the SEM image of the 15 nm thick Pb film shown in
Fig. \ref{cap:1}. The circles enclose clusters of grains that are
presumed to be centered on holes in the substrate.\label{cap:5}}
\end{figure}

Close inspection of the Pb SEM images suggests that this free energy
difference provides the force driving the grains to form over the
holes. In the coalescence process, grains hanging over a hole find
it energetically favorable to pull their partner toward the hole to
minimize the surface in contact with the substrate. This type of process
`roughens' the edges of the holes on the 80 nm hole substrate at $d=10$
nm and leads to the near complete closing of the holes at $d=20$
nm. On the 34 nm hole substrate (see Fig. \ref{cap:5}), it is possible
to see clusters of grains centered on a point where a hole is likely
to be. The regular arrangement of the larger grains around them makes
it possible to ascertain the positions of the clusters relative to
underlying holes. The grains in these clusters presumably are on the
verge of coalescing and would have coalesced and left a large grain
over the hole, had the film been made slightly thicker.

Within this simple picture it is possible to understand why the Pd,
Ge and Au films do not form ordered arrays of grains. In the case
of the Pd and Ge films, adatom diffusion rates are so low at room
temperature that the film structure is consistent with the deposited
atoms sticking where they land and wetting the substrate. Presumably
the interactions between the substrate and these elements are very
strong. Grain formation in the Au films suggests that there is some
adatom motion. The grain size, however, depends very weakly on film
thickness indicating that grain coalescence does not occur in Au films
on room temperature substrates. Typically, $T_{s}/T_{m}>0.24$ is
necessary for the thermally activated motion necessary for grain growth
through coalescence \cite{9}. For $T_{s}=300$ K, $T_{s}/T_{m}$
is 0.22 for Au.

It is important to note that other methods that produce nano-dots
with AAO substrates differ qualitatively from the one presented in
this paper. Masuda and coworkers separated the anodized portion of
the substrate from the underlying aluminum and used it as an evaporation
mask to produce arrays of metal dots \cite{10}. Gao and coworkers
produced an ordered array of hemispherical Au dots by depositing Au
on the underlying aluminum portion. This substrate had hemispherical
depressions in its surface that the evaporated metal filled. It is
likely that surface energies played a role in driving the Au into
the depressions.

The mechanism that we have considered suggests that it should be possible
to extend this technique to the formation of arrays of other metals
or nanostructures of other geometries. For instance, we speculate
that higher melting temperature non-wetting metals such as Au would
also show grain ordering if the deposition is carried out at higher
substrate temperatures. The spontaneous formation of these arrays
through the interaction of the deposited films with the substrate
structure suggests that it may be possible to engineer other nano-structures
in thin films using suitably patterned substrates. For example, one
could envisage the spontaneous formation of an array of parallel metal
wires on a substrate with narrow parallel trenches.

Potentially, the fabrication of these grain arrays can be used as
the first step in the creation of ordered superconducting junction
arrays consisting of nanometer scale superconducting islands. An overlayer
of semiconductor or metal deposited on the arrays could provide the
necessary inter-island coupling. The lack of processing involved in
their fabrication ensures that the interfaces in the junctions remain
free of contaminants. Moreover, the small size of these islands and
the spatial extent of their arrays are hard or impossible to produce
using standard lithographic techniques. Thus, it seems plausible to
create novel ordered arrays with a superconductor to insulator that
can be compared to that of disordered granular films\cite{11}. Similarly,
these arrays are suitable for investigations of a recently proposed
superconductor to metal quantum phase transition \cite{12,12too,12three}.

Pb and Sn films deposited on room temperature AAO substrates with
a hexagonal array of holes spontaneously form an ordered array of
nanometer scale crystals. The arrays first form at film thicknesses
at which the grain sizes are comparable to the hole diameters. The
grains appear to grow primarily through coalescence. Since the films
do not wet the substrate at all, the coalescence events are biased
by surface energies to drive the grains over the holes. Films composed
of grains that do not coalesce (e.g. Au) or that wet the substrate,
do not form ordered arrays of grains (e.g. Pd, Ge).

Acknowledgements: We are grateful for helpful discussions with Professors
Ben Freund and Humphrey Maris. This work has been supported by the
NSF through DMR-0203608 and an REU supplement, AFRL and ONR.

\bibliography{nprefs}

\end{document}